\documentclass[11pt,a4paper]{article}
\usepackage{jheppub}
\usepackage{graphicx}

\def\hbar{\hspace{0pt}\raisebox{1pt}{$-$} \hspace{-7pt} h}

\def\5{\overline 5}

\newcommand{\ba}{\begin{eqnarray}}
\newcommand{\ea}{\end{eqnarray}}
\newcommand{\no}{\nonumber}
\newcommand{\be}{\begin{equation}}
\newcommand{\ee}{\end{equation}}
\newcommand{\bea}{\begin{eqnarray}}
\newcommand{\eea}{\end{eqnarray}}

\title{A Natural Hierarchy and a low New Physics scale from a Bulk Higgs}
\date{\today
}
\author{
Luca Vecchi}
\affiliation{Theoretical Division T-2, Los Alamos National Laboratory \\ Los Alamos, NM 87545, USA}
\emailAdd{vecchi@lanl.gov}
\abstract{

We show that a bulk Higgs with a mass saturating the Breitenlohner-Freedman bound can naturally generate and stabilize an exponential hierarchy on a nearly AdS background. The physical Higgs boson in this class of models emerges as the lightest eigenstate of the Higgs/radion system and has a mass strictly lighter than the Kaluza-Klein scale. These theories are dual to strongly coupled CFTs deformed by a marginally relevant Higgs mass operator. On the 5D side, the marginally relevant nature of the Higgs mass operator implies that the Higgs VEV is maximally spread in the bulk. This feature significantly decreases the lower bound on the new physics scale in models that address the SM flavor problem. The collider phenomenology interpolates between Randall-Sundrum scenarios with a heavy Higgs and a light radion, and composite Higgs models.

}

\begin{document}
\maketitle

\section{Introduction}

Warped extra dimensional theories~\cite{RS} are among the most compelling models for physics beyond the standard model. These theories offer an elegant solution to both the hierarchy problem and the flavor puzzle, and predict a rich collider phenomenology.

In the original Randall-Sundrum (RS) model~\cite{RS} the 5D coordinate is confined between two boundaries, usually referred to as UV and IR branes. The local $AdS_5$ geometry generates an exponential relation between the fundamental parameters of the theory -- naturally of the order of the cutoff scale $\Lambda$ -- and the IR scale $\Lambda_\chi$, and therefore elegantly explains the hierarchy $\Lambda_\chi\ll\Lambda$ as a consequence of gravitational redshift.

In the absence of a stabilizing mechanism, the distance between the UV and IR boundaries is associated to a massless modulus, the radion, and it is therefore undetermined. Any realistic realization of the scenario should explain why the radion is massive, i.e. why the "interbrane" distance is stabilized. The simplest stabilizing mechanism~\cite{GW} consists in triggering a slight deformation of the $AdS_5$ geometry with the non-trivial profile of a bulk scalar field $\phi$. If the 5D mass $m^2$ of the scalar is small compared to the $AdS_5$ curvature scale $\Lambda$, say $(m/\Lambda)^2=O(0.1)$, the relation $\Lambda_\chi\ll\Lambda$ is naturally explained without fine-tuning, and the radion acquires a small mass $m_\chi^2=O((m/\Lambda)^c)\Lambda_\chi^2$ with $c>0$ a model-dependent number.

The dual field theory interpretation of this mechanism has been analyzed in~\cite{RZ}. The RS model without stabilizing field is dual to a 4D conformal field theory (CFT) in which a scalar operator with a large scaling dimension has acquired a vacuum expectation value (VEV). Due to the large scaling dimension, spontaneous conformal symmetry breaking appears on the 5D side as a sharp IR cutoff (the IR boundary). The massless radion is then interpreted as a dilaton, the Nambu-Goldstone boson associated to the spontaneous CFT breaking. 

The dual viewpoint also offers a nice perspective on the physics of the stabilization mechanism. In order to stabilize the hierarchical relation between the IR and UV scales, and thereby provide a tiny mass for the dilaton, one should introduce a small, explicit deformation of the CFT. This is accomplished by adding to the CFT an almost marginal local operator. 

The simplest CFT deformation has the form
\ba\label{def1}
\delta{\cal L}_{\textrm\tiny{{GW}}}=J{\cal O}.
\ea
If the operator ${\cal O}$ has a scaling dimension $\Delta\approx4$ the explicit CFT breaking can be considered "small". The net effect of~(\ref{def1}) is to generate a slowly varying potential $V_{eff}=\chi^4P((\chi/\Lambda)^{|\Delta-4|})$ for the dilaton $\chi$. The vacuum $\langle\chi\rangle\equiv\Lambda_\chi$ associated to the mass gap, is now determined dynamically. A huge hierarchy $\Lambda_\chi\ll\Lambda$ and a small mass $m_\chi$, typically proportional to the explicit breaking parameter, are then generated without fine-tuning. The choice~(\ref{def1}) is precisely the one made in~\cite{GW}.

The next to simplest deformation has the form
\ba\label{def2}
\delta{\cal L}_{\textrm\tiny{{TCC}}}=\lambda{\cal O}^\dagger{\cal O}.
\ea
Now the explicit CFT breaking can be considered "small" if the operator ${\cal O}$ has a scaling dimension $\Delta\approx2$. The physics associated to the CFT deformation~(\ref{def2}) has been recently studied using field theory and gauge/gravity techniques in~\cite{Vecchi} and~\cite{multi}. An explicit field theory model for dynamical symmetry breaking based on the deformation~(\ref{def2}) has been analyzed in~\cite{TCC}. For reasons explained in the latter reference we will refer to the class of CFTs deformed by~(\ref{def2}) as Technicolor at Criticality (TCC).

The IR relevance of the deformation~(\ref{def2}) is mainly controlled by quantum effects. In order to appreciate this, let us assume $\Delta=2$ for simplicity~\footnote{The following discussion is not altered as long as the corrections to the condition $\Delta=2$ (for example from subleading orders in $1/N$, see~\cite{TCC}) are somewhat smaller than $\lambda$. See Appendix~\ref{A} for details.}. In this case, the coupling $\lambda$ in~(\ref{def2}) is classically marginal. However, as soon as quantum corrections are taken into account the beta function $\beta_\lambda=\mu\frac{d\lambda}{d\mu}$ develops a quadratic contribution
\ba\label{betaL}
\beta_\lambda=-\lambda^2
\ea
which tells us that for $\lambda>0$ the coupling is actually \textit{marginally relevant}. If the CFT admits a large N expansion suppressing higher order correlators of ${\cal O}$, the beta function~(\ref{betaL}) is exact in the planar limit~\cite{Witten}. 

Despite the fact that $\lambda$ grows fast in the IR, the explicit (UV), hard CFT deformation is "small" (almost marginal indeed). The very same arguments applied to the~(\ref{def1}) case say that~(\ref{def2}) induces a potential $V_{eff}=\chi^4P(\log\chi/\Lambda)$ for the dilaton $\chi$, and hence triggers the generation of a huge hierarchy $\langle\log\chi/\Lambda\rangle=O(1)$ between fundamental and dynamical scales. Because there is no parameter controlling the smallness of the deformation at the dynamical scale $\langle\chi\rangle=\Lambda_\chi$, the dilaton mass is generally expected to be $m_\chi=O(\Lambda_\chi)$ in the TCC scenario.


A theoretical advantage of the choice~(\ref{def2}) over~(\ref{def1}) is that in the case of~(\ref{def2}) the operator ${\cal O}$ [and as a consequence its dual 5D field $\phi$] is allowed to carry a global charge. We propose to identify ${\cal O}$ in~(\ref{def2}) with the Higgs operator. With such an identification the Higgs field would be responsible for both breaking the electro-weak symmetry \textit{and} stabilizing the hierarchy, and the role of the dilaton $\chi$ would be played by the Higgs boson itself. Let us now elaborate on this latter point further.

The \textit{chiral} symmetry, under which ${\cal O}$ is charged, is spontaneously broken by the non-trivial profile of the dual 5D scalar $\phi$. This phenomenon is expected to occur when the coupling $\lambda$ becomes non-perturbative. Taking advantage of the exact result~(\ref{betaL}), and using the fact that the expectation value of the conformal anomaly of our theory, i.e. $\langle\beta_\lambda{\cal O}^\dagger{\cal O}\rangle$, is RG invariant, it follows that the running scaling dimension of the order parameter ${\cal O}$ in TCC reads~\cite{Vecchi}
\ba\label{dim}
\Delta[{\cal O}]=2-\lambda.
\ea
The regime in which the non-perturbative effects are expected to turn on, say when $\lambda\sim1$, coincides with the regime in which the scaling dimension of the order parameter ${\cal O}$ gets close to the dimension $\Delta[{\cal O}]=1$ of a free scalar [i.e. of a pseudo-dilaton !]. A possible interpretation of this [otherwise scheme-dependent] result is that TCC has in fact a remnant of conformal invariance in the IR, and that the Higgs boson in such a scenario \textit{is} a pseudo-dilaton. In this sense the composite Higgs boson of the TCC scenario is \textit{weakly coupled}~\cite{TCC}.

The aim of the present paper is to formulate the TCC framework on a weakly coupled 5D world in an effort to gain a quantitative understanding of this scenario.

\section{The Model}

In this section we show how to reproduce the field theory~(\ref{def2}) on a 5D set up~\footnote{See also~\cite{FHR} for an application to condensed matter physics.}. The starting point is clearly an exact CFT, which on the gravity side is realized as an $AdS_5$ background:
\ba\label{AdS}
ds^2=\frac{1}{(\Lambda z)^{2}}(\eta_{\mu\nu}dx^\mu dx^\nu-dz^2).
\ea 
There are now two basic ingredients we need in order to realize our program. First, we need a bulk Higgs charged as a $(2,1/2)$ of the standard model $SU(2)_L\times U(1)_Y$ gauge group. We write the Higgs doublet as $U(0\;\phi)^t$, with $U$ an unitary matrix defined in terms of the 3 unphysical Nambu-Goldstone modes eaten by the $W^\pm$ and $Z^0$ after electro-weak symmetry breaking, and $\phi$ the field acquiring a nontrivial vacuum. In the following we will focus on the electromagnetic singlet scalar $\phi$ which, with an abuse of notation, will be called the Higgs operator. We assume the bulk Higgs has a 5D mass saturating (or close to) the Breitenlohner-Freedman bound~\cite{BF}
\ba\label{m}
m^2=-4\Lambda^2.
\ea
The field thus defined is dual to the dimension $\Delta=2+\sqrt{(m/\Lambda)^2+4}=2$ Higgs operator ${\cal O}$ of the field theory. In Appendix~\ref{A} we will show that small deviations from the condition $\Delta=2$ do not alter our discussion. The second key ingredient consists in the implementation of the deformation~(\ref{def2}). Let us discuss this latter point in some detail.

The condition~(\ref{m}) implies that the asymptotic expression for the scalar field close to the UV region $z\sim1/\Lambda$ is 
\ba\label{asympt}
\phi=z^2(\alpha\log\mu z+\beta), 
\ea
with $\alpha,\beta$ generally functions of the 4D coordinates and $\mu<\Lambda$ an RG scale. The AdS/CFT correspondence [for $\Delta\geq2$] associates $\beta$ to the VEV of the dual operator ${\cal O}\leftrightarrow\phi$ [typically in the presence of CFT deformations] whereas $\alpha$ [typically a function of $\beta$] to the CFT deformation. Specifically, given a CFT deformation of the generic form $\delta{\cal L}=F({\cal O})$ the relation between the coefficients $\alpha,\beta$ turns out to be $\alpha=F'(\beta)$~\cite{Witten}. For example, from a linear deformation $F=J{\cal O}$, see~(\ref{def1}), one recovers the conventional identification $\alpha=J$. In order to reproduce the TCC formalism, on the other hand, we should require $\alpha\propto\beta$: the implementation of the deformation~(\ref{def2}) on a 5D set-up requires that both coefficients $\alpha,\beta$ in~(\ref{asympt}) are non-vanishing. This is perhaps the reason why the TCC scenario has not received much attention so far.

To implement the relation $\alpha\propto\beta$, it is convenient to introduce an UV boundary at $z=1/\mu$ and an UV boundary potential for the Higgs. The UV potential is:
\ba\label{BC}
V_{UV}=\left(2-\lambda\right)\Lambda\,|\phi|^2+\dots,
\ea
where the dots indicate possible constant or irrelevant terms in $\phi$. The potential~(\ref{BC}) is used to enforce the UV condition $z\phi'=(2-\lambda)\phi$ from which the relation
\ba\label{BCA'}
\alpha=-\lambda\beta
\ea
follows. Here, $\lambda=\lambda(\mu)$ plays the role of the renormalized coupling of the deformation~(\ref{def2}) evaluated at the scale $\mu$, and it has been defined so that for $\lambda(\mu)>0$ the deformation is marginally relevant, as we will see. A systematic implementation of CFT deformations along these lines has been presented in~\cite{multi}.

The bulk Higgs profile is completely specified once the bulk potential and the IR boundary condition are given. For computational reasons, in the following we will work at leading order in the backreaction [we will comment later on the possibility of generalizing our results], and therefore we limit our study to quadratic order in $\phi$. In this case the 5D action for the scalar simply reads
\ba\label{I}
{\cal S}_\phi&=&\int d^4x\int_{1/\mu}^{1/\chi} dz\sqrt{-g}\bigg\{|\nabla\phi|^2-m^2|\phi|^2+\dots\bigg\}\\\no
&+&\int_{z=1/\mu} d^4x\sqrt{-g_{UV}}\left[-V_{UV}\right]\\\no
&+&\int_{z=1/\chi} d^4x\sqrt{-g_{IR}}\left[-V_{IR}\right],
\ea
where $z=1/\chi$ and $z=1/\mu$ define the location of the IR and UV boundaries, respectively, and $\nabla$ is the gauge covariant derivative~\footnote{We do not specify the gauge symmetry in the bulk because highly model-dependent. This is at least the $SU(2)\times U(1)$ symmetry of the standard model, but it could be larger if a custodial symmetry is assumed.}. The dots refer to higher order terms in $\phi$, the cosmological constant, and terms involving other bulk fields such as gauge bosons or fermions; $V_{UV}$ is given in~(\ref{BC}), whereas for definiteness the potential $V_{IR}$ is chosen such as to fix $\phi(z=1/\chi)=\phi_*$. 

It is import to stress that the explicit form of the bulk and IR potentials is not relevant for our discussion: the key ingredients required to reproduce the TCC scenario on a 5D world are~(\ref{m}) and~(\ref{BCA'}). Our ignorance regarding the unspecified terms appearing in~(\ref{I}) will be effectively parametrized by two parameters $a,b$ in the following section.

Within our approximation the bulk Higgs profile reads
\ba\label{profile}
\phi=\phi_*\left(\chi z\right)^2\frac{1-\lambda \log\mu z}{1-\lambda \log\mu/\chi}.
\ea
Because we are identifying $\phi$ with the Higgs field, the quantity $\phi_*$ is determined by the 4D electro-weak vacuum $v=250$ GeV. At leading order in $v$, the mass of the $W^\pm$ boson is given by
\ba\label{mW}
m_W^2=\frac{g_5^2}{2}\frac{\int dz \left(\frac{1}{\Lambda z}\right)^3 \phi^2}{\int dz \left(\frac{1}{\Lambda z}\right)}\equiv\frac{g^2 v^2}{4},
\ea
with $g_5$ the dimensionful 5D gauge coupling. By conventionally identifying the 4D gauge coupling with $g^2=g_5^2/\int dz \left(\frac{1}{\Lambda z}\right)$, we interpret~(\ref{mW}) as the \textit{definition} of $\phi_*$.

Before closing this introductory section we mention that the theory also contains a purely gravitational term, which in the absence of ${\cal S}_\phi$ reproduces the $AdS_5$ background~(\ref{AdS}) [our CFT]. The gravity action is
\ba\label{grav}
{\cal S}_{\small{\textrm{gravity}}}&=&\int d^4x\int_{1/\mu}^{1/\chi} dz\sqrt{-g}\bigg\{-\frac{M^3}{2}R+6M^3\Lambda^2\bigg\}\\\no
&+&\int_{z=1/\mu} d^4x\sqrt{-g_{UV}}\left[M^3K-3M^3\Lambda^2\right]\\\no
&+&\int_{z=1/\chi} d^4x\sqrt{-g_{IR}}\left[M^3K+3M^3\Lambda^2\right].
\ea
Here $K$ is the trace of the extrinsic curvature and allows us to formulate a consistent variational problem for gravity in a space with boundaries. Our model is finally described by the 5D action
\ba\label{tot}
{\cal S}={\cal S}_{\small{\textrm{gravity}}}+{\cal S}_\phi,
\ea
where all dimensionful quantities are assumed to be at the natural scale $\Lambda$, namely $M,\phi_*^{2/3}$, and $g_5^{-2}$ are all $O(\Lambda)$. The bare parameters $g_5,\phi_*,\lambda$, and $M$ will be eventually traded for the physical quantities $g$ [the 4D gauge coupling], $v$ [the electro-weak vacuum], $\Lambda_\chi$ [the dynamical scale], and $f$ [the dilaton decay constant].

In the absence of ${\cal S}_\phi$ the theory relaxes to the $AdS_5$ vacuum~(\ref{AdS}) with $\chi\rightarrow0$. We will see shortly that as soon as the deformation~(\ref{def2}) is switched on [i.e. as soon as $\lambda\neq0$ in~(\ref{profile})], the theory picks up another vacuum with $\chi=\Lambda_\chi\ll\Lambda$.

\subsection{The dilaton effective action}

In order to determine the vacuum of the theory~(\ref{tot}) we follow~\cite{GW}~\footnote{See Appendix~\ref{A'} for a rigorous, non-perturbative derivation of the vacuum configuration of a generic scalar/gravity system with boundary conditions~(\ref{BCA'}).} and identify $\chi$ with a 4D dynamical field (the dilaton); we then derive the effective 4D action for $\chi$ by evaluating the action~(\ref{tot}) with the line element~(\ref{AdS}) and the scalar $\phi$ determined by its (classical) equations of motion, see~(\ref{profile}). Focusing on the leading contributions in $\phi^2_*/\Lambda^3$ we have
\ba\label{Ldil}
{\cal S}=\int d^4x \left[\frac{3}{2}\frac{M^3}{\Lambda^3}(\partial\chi)^2-V_{eff}\right],
\ea
with the kinetic term following from~(\ref{grav}), while $V_{eff}$ entirely from~(\ref{I}). At leading order the expression for the \textit{renormalized} effective potential at the RG scale $\mu$ is very simple (see also Appendix~\ref{A}):
\ba\label{Veff}
V_{eff}&=&\frac{1}{(\Lambda z)^3}\left[\phi\phi'\right]_{z=1/\chi}+\textrm{C.C. terms}\\\no
&=&\frac{\phi_*^2}{\Lambda^3}\left[\chi^4\left(a-\frac{\lambda(\mu)}{\lambda(\mu)\log\chi/\mu+1}\right)+b\Lambda^{4}\right],
\ea
where "$\textrm{C.C. terms}$" refers to the cosmological constant terms coming from the bulk and boundary potentials. As anticipated, we parametrized these contributions with the constants $a,b$.

The quantities $a,b$ depend on the explicit expression of the scalar potentials and are therefore model-dependent. Because $b$ simply reduces to a constant in $V_{eff}$, it plays no role in the determination of the vacuum solution. Its role is to ensure the vanishing of the effective potential on the vacuum solution, and its actual value is related to the well known cosmological constant problem. We will not address this issue in the following and we will hence focus on the non-trivial part of~(\ref{Veff}). 

The constant $a$ will be left essentially arbitrary in what follows, except for the requirement $a>0$. The reason why we impose the latter constraint is that the field theory discussed in the Introduction assumes that in the absence of the deformation~(\ref{def2}) the theory is an exact CFT. If our 5D framework has to reproduce this physics, the true vacuum for $\lambda=0$ should be given by $\langle\chi\rangle=0$. An inspection of~(\ref{Veff}) reveals that this happens only if $a>0$, otherwise the undeformed theory would be unstable ($a<0$) or it would manifest spontaneous CFT breaking ($a=0$).

We are now ready to analyze the physics emerging from of the effective potential~(\ref{Veff}). First of all we see that the potential has the form anticipated in the Introduction, namely $V_{eff}=\chi^4P(\log\chi)$. This fact follows from~(\ref{m}) and the UV condition~(\ref{BCA'}). One can also check with the help of~(\ref{betaL}) that~(\ref{Veff}) consistently satisfies the Callan-Symanzik equation, ensuring that the dependence of $V_{eff}$ on the RG scale $\mu$ is not physical. Furthermore, in the case $\lambda>0$ of our interest [for $\lambda<0$ the theory is IR free] the absolute minimum of the potential sits at 
\ba\label{vacuum}
\langle\chi\rangle\equiv\Lambda_\chi=\Lambda e^{-\frac{1}{\lambda(\Lambda)}+\frac{1}{\lambda_*}},\quad\quad \lambda_*\equiv \lambda(\Lambda_\chi)=\frac{2a}{1+\sqrt{1-a}}.
\ea
Eq.~(\ref{vacuum}) is the solution of the quadratic beta function~(\ref{betaL}). This confirms that our 5D scenario correctly implements the main feature of the TCC framework. The scale $\Lambda_\chi$ sets the mass gap in the problem, with a typical Kaluza-Klein (KK) mass given by
\ba
m_\rho=2.4\,\Lambda_\chi.
\ea
The vacuum solution~(\ref{vacuum}) only exists in the range $0<a\leq1$, with the constraint $a>0$ ensuring that the CFT deformation is responsible for CFT breaking, i.e. for the IR brane generation. In the following we will thus assume that
\ba\label{range}
0<\lambda_*\leq2.
\ea 
Field theory arguments also support the guess $\lambda_*=O(1)$.

For $\lambda_*>\lambda$ the CFT deformation~(\ref{BC}) is marginally relevant and our model manifests spontaneous compactification at a large length scale $\Lambda_\chi\ll\Lambda$. As shown in Appendix~\ref{A} such a hierarchical relation is not specific to~(\ref{m}). Instead, it is a generic implication of~(\ref{BC}) and is found whenever the CFT deformation is small at the cutoff scale, i.e. whenever
\ba\label{natural}
2\sqrt{\frac{m^2}{\Lambda^2}+4}<\lambda(\Lambda)<1.
\ea
In our framework the relation~(\ref{natural}) is not spoiled by radiative corrections as long as $\lambda(\Lambda)$ is not too small, say $\lambda(\Lambda)\gtrsim1/4\pi$. Absence of fine-tuning in our model therefore requires new physics to enter not far above $10^6\Lambda_\chi$.

\subsection{The Higgs boson}

Our approach so far has been rather heuristic: we assumed that $\chi$ is a dynamical variable and did not take into account the backreaction of the profile~(\ref{profile}) on the geometry [see however Appendix~\ref{A'}]. The first assumption is actually a very good approximation: despite the fact the $\chi$ does not satisfy the linearized equations of motion of the scalar/gravity system, the scalar $\chi$ can be thought of as the radion~\cite{dilaton}, i.e. as the dilaton of the dual theory. The second assumption deserves an explanation.

The physical scalar excitations emerging from our model correspond to a mixture of the KK tower of the bulk scalar $\phi$ and the radion, the scalar excitation of the 5D metric. The lightest mass eigenstate resulting from the diagonalization of this system is to be identified with the four dimensional Higgs boson.

Our results have been evaluated under the assumption that $\phi^2/M^3\ll1$, namely under the assumption that the backreaction of the scalar on the metric is small. Yet, numerical simulations of the full backreacted system show that the physics of the lightest scalar eigenstate -- our Higgs boson -- is remarkably well approximated by a leading approximation, like the one performed here, even for $\phi^2/M^3=O(1)$ or bigger~\cite{Peloso}\cite{PE}. In particular, it turns out that the mass of the lightest scalar is given to a good approximation by
\ba\label{V''}
m_\chi^2&=&\frac{\Lambda^3}{3M^3}V''_{eff}\\\no
&=&v^2\frac{\Lambda^3}{3M^3}\frac{2\lambda_*^2(2-\lambda_*)}{1+\lambda_*+\lambda_*^2/2},
\ea
and that its couplings are controlled to a very high accuracy by the \textit{decay constant} $f$, defined as the vacuum expectation value of the canonically normalized dilaton field~\footnote{The canonically normalized dilaton $\bar\chi$ is defined by $\bar\chi=\sqrt{3(M/\Lambda)^3}\chi$ (recall that in the present formalism $M$ is not related to the 4D Planck mass).}:
\ba\label{fdil}
f^2\equiv3\left(\frac{M}{\Lambda}\right)^3\,\Lambda_\chi^2.
\ea
The physics of the lightest mode is correctly described by the dilaton $\chi$ even for large $\phi^2/M^3$ because the actual mixing Higgs/dilaton is negligible as long as the lightest mode is parametrically lighter than the KK excitations (i.e. than the composite Higgs). If this latter condition is fulfilled then the lightest eigenstate is almost dominantly $\chi$, irrespective of the magnitude of the backreaction. We conclude that, as long as $m_\chi^2\ll m_\rho^2$, the analysis performed in this section is reliable even for $\phi^2/M^3=O(1)$, and that the Higgs boson in our model behaves as a dilaton field $\chi$ with decay constant $f$~\cite{TCC}.


In fact, it is unnatural to assume a hierarchical relation between $\phi^{2/3}$ and the 5D Planck scale~\footnote{In~\cite{TCC} both quantities are of $O(N^2)$ and one anticipates $\phi_*^2/M^3=O(1)$.}, as both quantities are expected to be of $O(\Lambda)$. Using~(\ref{mW}) and~(\ref{vacuum}) we have
\ba
v^2=\frac{\phi_*^2}{\Lambda^3}\left(1+\lambda_*+\lambda_*^2/2\right)\Lambda_\chi^2,
\ea
that says that when the natural relation $\phi_*^2\approx M^3$ is satisfied one finds $f\approx v$: the dilaton decay constant approaches the electro-weak vacuum and the (renormalizable) dilaton couplings (the couplings of our Higgs boson) become formally the same as those of a classical Higgs boson~\cite{GGS}\cite{dilaton}! More generally one should require $v\lesssim f$ for the semiclassical approach to gravity to be sensible.

As explained above, the ratio $m_\chi^2/m_\rho^2$ controls the reliability of our approach, and we would like it to be small. Consistently, we find that the Higgs/dilaton mass is always suppressed, even in the absence of small parameters, namely:
\ba
m_\chi^2\equiv c \frac{v^2}{f^2}m_\rho^2\quad\quad\quad0\leq c(\lambda_*)\lesssim\frac{1}{10},
\ea
where the explicit form of $c$ can be read from eqs.~(\ref{V''}) and~(\ref{fdil}), and the bound is a consequence of the consistency condition~(\ref{range}). Recalling that the parameter $\lambda_*$ is controlled by the (unspecified) bulk and IR potentials (via $a$), we see that $c$ is an intrinsic property of the strong dual dynamics and that, in particular, there is no UV parameter that can affect $m_\chi^2$. In this sense the guess $m_\chi=O(\Lambda_\chi)$ given in the introduction is confirmed~\cite{TCC}.

\begin{figure}
\begin{center}
\includegraphics[width=4.5in]{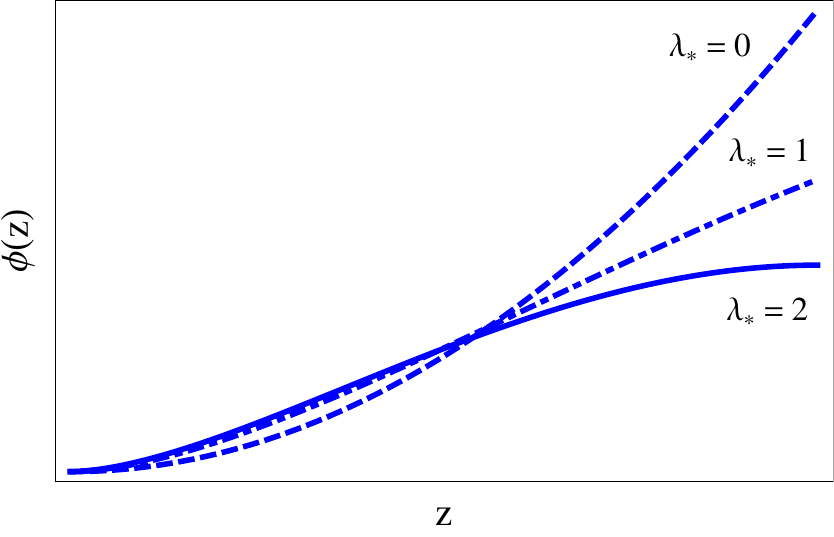}
\caption{\small Higgs VEV~(\ref{HVEV}) for various values of the coupling $\lambda_*$ and $1/\Lambda\leq z\leq1/\Lambda_\chi$. The VEV is normalized in such a way that for any $\lambda_*$ it reproduces the electro-weak scale via~(\ref{mW}). \label{fHVEV}}
\end{center}
\end{figure}

One can show that the dilaton mass has precisely the structure 
\ba
m_\chi^2=-4\beta_\lambda\frac{\langle{\cal O}^\dagger{\cal O}\rangle}{f^2}
\ea
dictated by conformal invariance. As expected, the dilaton mass is suppressed if $\lambda_*\ll1$ [this limit is not of interest to us because it appears as an unnatural realization on the field theory side]. More interestingly, though, we find that $m_\chi^2$ is also suppressed if $\lambda_*$ is close to the critical value $\lambda_*=2$ ($a=1$), at which $V'_{eff}$ develops a double zero~\footnote{The existence of a critical value for the IR brane tension for which the dilaton mass vanishes is a generic property. The remarkable feature typical of the present model is that -- even though the model itself exists for a limited range of $a$, see~(\ref{range}) -- such a critical value is in fact allowed.}. A look at the scalar profile~(\ref{profile}), which thanks to the definitions~(\ref{mW}) and~(\ref{vacuum}) can be re-written as~\footnote{For $\lambda_*=0$ this expression agrees with~\cite{gaugephobic}.}
\ba\label{HVEV}
\phi=\frac{v}{\Lambda_\chi}\Lambda^{3/2}\left(z\Lambda_\chi\right)^2\frac{1-\lambda_*\log z\Lambda_\chi}{\sqrt{1+\lambda_*+\lambda_*^2/2}},
\ea
reveals that this condition is met when the Higgs VEV becomes flat in the IR, i.e. when $\phi'(z=1/\Lambda_\chi)=0$, see fig.\ref{fHVEV}: in this case a remnant of conformal invariance (explicitly broken by the RG flow of $\lambda$) is non-linearly realized in the field theory, and the dilaton mass is anomalously suppressed.

It would be interesting to find an explicit solution of the nonlinear gravity/scalar system  (see Appendix~\ref{A'}) and identify which properties of the scalar potential govern the condition $\lambda_*\sim2$, and ultimately which properties of the strong dynamics are required in order to generate a light Higgs boson in the spectrum.

\section{Experimental constraints}

The most stringent phenomenological bounds in our scenario come from precision electro-weak measurements and flavor constraints, and depend on whether the standard model (SM) fermions are fundamental with respect to the strong, bulk dynamics or partially composite. In the former case the SM fermions are localized on the UV boundary, while in the latter case they propagate in the bulk.

\subsection{Electro-weak precision tests}

The LEP experiment has put severe bounds on the electro-weak (EW) precision parameters $S$ and $T$. Here we will be mainly concerned on the $S$ parameter, as $T$ can be controlled by imposing (or explicitly breaking) a bulk $SU(2)$ custodial symmetry~\cite{ADMS}. The remaining EW precision parameters are generally suppressed in warped models and will be ignored.

Let us thus focus on the $S$ parameter and assume that \textit{all} SM fermions are localized close to the UV boundary. In this case an explicit (tree-level) expression for $S$ can be written as
\ba\label{S}
S=\frac{8\pi}{g_5^2\Lambda}\int \frac{dz}{z}\left(1-\rho^2\right),
\ea
where the function $\rho$ satisfies the massless differential equation for the $W^\pm$, with the UV boundary condition $\rho(z=1/\Lambda)=1$. In the presence of a bulk Higgs $\rho$ is defined by 
\ba
-\rho''+\frac{1}{z}\rho'+\frac{1}{2}\left(g_5\frac{1}{\Lambda z}\phi\right)^2\rho=0,
\ea
and it is subject to the IR boundary condition $\rho'(z=1/\Lambda_\chi)=0$. Working at leading order in $m_{W}^2$ and for a Higgs VEV given by~(\ref{HVEV}) we find
\ba\label{STCC}
S=\frac{3\pi}{4}\left(\frac{1+\frac{2}{3}\lambda_*+\frac{5}{24}\lambda_*^2}{1+\lambda_*+\frac{1}{2}\lambda_*^2}\right)\frac{v^2}{\Lambda_\chi^2}.
\ea
For $\lambda_*=0$ the $S$ parameter has been computed by several authors (see for example~\cite{Piai} and~\cite{Yam}). The result~(\ref{STCC}) is in agreement with these studies. For comparison, we also notice that for a Higgs infinitely localized in the IR, and following the very same procedure but now with $\phi=0$ and the IR condition $\rho'\propto\rho$, one gets $S=\pi v^2/\Lambda_\chi^2$. 


For any $\lambda_*>0$ compatible with the constraint~(\ref{range}) we observe a suppression of the $S$ parameter compared to the case $\lambda_*=0$ [and hence to the IR Higgs case]. The suppression increases with $\lambda_*$ and it is maximal for $\lambda_*=2$. This effect is a consequence of the decreasing overlap of the Higgs VEV~(\ref{HVEV}) with the IR (see fig.\ref{fHVEV}). Assuming a light Higgs [$m_\chi\sim110$ GeV] we require $S\lesssim0.35$ at $99\%$ CL~\cite{Barbieri} and find $\Lambda_\chi\gtrsim870\,(1100)$ GeV, or equivalently
\ba\label{rel}
m_\rho\gtrsim2.1\,(2.6)\,\textrm{TeV},
\ea
for $\lambda_*=2(0)$ [one should also arrange $T\lesssim0.35$]. The relatively low mass scale~(\ref{rel}) is comparable to that found in holographic composite Higgs models~\cite{CH}.

If the SM fermions propagate in the bulk an additional contribution to $S,T$, typically of the same order as the purely oblique estimate given in~(\ref{STCC}), is present. We will not discuss this correction [nor corrections to the $Z^0\rightarrow b\bar b$ coupling] because in such a scenario flavor bounds, on which we now turn, are typically stronger~\cite{Casagrande}.

\subsection{Flavor Physics}

The flavor physics in an RS background with bulk fermions can be understood as follows. In the dual language fundamental fermions $\psi_i$ ($i$ is a flavor index) mix with fermionic CFT operators $Q_i$ of scaling dimension $\Delta[Q_i]=2+c_i$. The couplings between the two sectors are defined at a large scale $\Lambda_F$, and are governed by the flavor-dependent constant $c_i$ and the ratio
\ba
\epsilon\equiv\frac{\Lambda_\chi}{\Lambda_F}.
\ea
After EW symmetry breaking, the effective theory at the IR scale contains the mixing matrix
\ba\label{Feff}
y_F\epsilon^{\Delta-1}\psi\psi+y_f\epsilon^{c-1/2}\psi Q+y_*QQ,
\ea
where $y_F,y_f,y_*$ are assumed to be $O(1)$ and flavor indeces have been suppressed. In the above expression we included a direct coupling between the fundamental fermions and a composite Higgs of scaling dimension $\Delta$. In the case of an IR-localized Higgs one has $\Delta\rightarrow\infty$ and this term drops, while in our scenario $\Delta\sim2$. In the general case, integrating out the heavy CFT operators leaves us with a mass matrix of the form
\ba\label{mf}
m_f\approx\frac{v}{\sqrt{2}} \left[y_F\epsilon^{\Delta-1}-\frac{y_f^2}{y_*}\epsilon^{2c-1}+\dots\right],
\ea
where we neglected subleading terms (and again suppressed the flavor indeces). For $2c<\Delta$ the first term in the mass matrix~(\ref{mf}) can be discarded. This is the case typically considered in the literature, in which the light fermion masses can be easily suppressed as compared to the dynamical scale without invoking any fine-tuning if $2c>1$. In the opposite limit, $2c>\Delta$ (always absent for an IR-localized Higgs model) the first term dominates and the fermion mass matrix is saturated by physics at the scale $\Lambda_F$. 

The low energy effective field theory also contains 4-fermion interactions that potentially contribute to FCNC effects. These latter are generated by two competing mechanisms. First, there is a short distance contribution generated at the flavor scale $\Lambda_F$; second, there is a long distance, indirect contribution coming from the mixing with the heavy CFT fermions. Taking both contributions into account one has (in the interaction basis)
\ba\label{4ferm}
C_{ijkl}\psi_i\psi_j\psi_k\psi_l\quad\quad\textrm{with}\quad\quad C\approx\frac{1}{\Lambda_\chi^2}\left[g_F^2\epsilon^2+g_*^2y_f^4\epsilon^{4c-2}\right],
\ea
where to simplify the expression we neglected the flavor indeces. If all numerical coefficients are $O(1)$ the second, flavorful term dominates as far as $c<1$, whereas the first, flavor blind becomes dominant if $c>1$ (this latter effect has been called UV-dominance in~\cite{Bauer}). In the absence of a GIM-like mechanism suppressing flavor violating terms in the matrix $g_F$, consistency with experiments typically requires $\Lambda_F>10^5$ TeV (this is true for any $\Delta$).

Given~(\ref{mf}) and~(\ref{4ferm}) one identifies four distinct limits. In the first $2c>\Delta$, $c>1$, and both the fermion mass matrix and flavor violation are controlled by short distance effects. This is what happens in the SM (where $\Delta=1$) and in extended technicolor models (where generally $\Delta>1$). In the limit $2c<\Delta$, $c<1$, on the other hand, both quantities are dominated by long distance effects. One can now explain the SM flavor hierarchy with $\epsilon\ll1$ and $c=O(1)$, and flavor violation is controlled by the so called RS-GIM mechanism~\cite{RSGIM}. This scenario can easily be realized in a RS background with SM fermions in the bulk and a large flavor scale. The third limit is defined by the condition $\Delta<2c<2$, and it is not particularly interesting because it does not address the flavor puzzle while FCNC effects are potentially problematic. The last scenario has $\Delta>2c>2$ and seems compelling, though. In this latter case the flavor hierarchy is addressed and flavor violation appears to be controlled by short distance effects. Notice, however, that this latter condition does actually apply only if \textit{all} the SM fermions have $c>1$: if at least one flavor has $c<1$ (this is always the case for the top quark) then the largest contribution to FCNC effects comes again from the flavorful term in~(\ref{4ferm}).

TCC with its flavor scale typically around $\Lambda_F\sim10^6$ TeV (i.e. $\epsilon\sim10^{-6}$) has generally $c\gtrsim1.5$ for the first quark generation, $c\lesssim1$ for the second and the bottom quark, and $c\sim0.5$ for the top. In Appendix~\ref{B} we illustrate with an example that -- as anticipated above -- this is not sufficient for the short distance term in~(\ref{4ferm}) to dominate: after rotating the fields to the mass basis one finds that the larger contribution to FCNC effects is in fact IR dominated. Yet, the bound on the new physics scale $\Lambda_\chi$ in this model is less severe than in an RS scenario with IR-localized Higgs, i.e. with no direct Higgs contribution to the mass matrix~\cite{Agashe}.

To appreciate this we follow~\cite{Gino} and define the mass matrix for the 4D fermion zero modes $\psi_i(x)$ as
\ba\label{eff}
{\cal L}_{flavor}^{4D}&=& \sum_{ij}y_{5D}^{ij}\left[\int dz\left(\frac{1}{\Lambda z}\right)^5\psi_0(c_i,z)\psi_0(c_j,z)\phi\right]\psi_i\psi_j \\\no
&=&\sum_{ij}y_*^{ij}r_{00}^H(c_i+c_j)\frac{v}{\sqrt{2}}\, f(c_i)f(c_j)\psi_i\psi_j,
\ea
where $i,j$ are flavor indeces, $y_*^{ij}$ is the 5D Yukawa matrix normalized in units of the cutoff scale, and $r_{00}^H$ is a measure of the overlap of the fermion zero modes with the Higgs VEV. The functions $f(c)$ appear in the definition of the fermion eigenfunctions $\psi_0(c,z)$ and read
\ba\label{f}
f(c)\equiv\sqrt{\frac{2c-1}{(1/\epsilon)^{2c-1}-1}},
\ea
with $c_i$ being the 5D fermion mass in units of the cutoff $\Lambda$~\footnote{The bulk fermion mass term has been defined so that $c_i>0.5$ implies localization near the UV boundary for both chiralities.}. Notice that in the present discussion the UV cutoff is identified with the physical scale $\Lambda_F$.

For an IR localized Higgs the definition~(\ref{eff}) applies~\footnote{\label{foot}We are tacitly assuming that the limit $\Delta\rightarrow\infty$ in which the bulk Higgs becomes effectively IR-localized is smooth [see~\cite{TH} for details].} if $r_{00}^H=1$, whereas in our model
\ba\label{r00H}
r_{00}^H(\bar c)=\frac{\sqrt{2}}{\sqrt{1+\lambda_*+\lambda_*^2/2}}\frac{1-\epsilon^{2-\bar c}}{2-\bar c}\left[1+\frac{\lambda_*}{2-\bar c}-\frac{\lambda_*\log1/\epsilon}{\left(1/\epsilon\right)^{2-\bar c}
-1}\right],
\ea
where we defined $\bar c=c_i+c_j$. With these definitions we readily see that the mass matrix has the form anticipated in~(\ref{mf}). In particular we observe the three distinct regimes
\ba\label{mf'}
m_f^{ij}&=& y_*^{ij}r_{00}^H(c_i+c_j) f(c_i)f(c_j)\frac{v}{\sqrt{2}}\\\no
&\propto&y_* v\left\{
\begin{array}{rll}
\left(1+\frac{\lambda_*}{2-\bar c}\right) & \textrm{\quad for $\bar c<1$},\\ 

\\

\epsilon^{\bar c-1}\left(1+\frac{\lambda_*}{2-\bar c}\right) & \textrm{\quad for $1<\bar c<2$},\\ 

\\

\lambda_*\epsilon\log 1/\epsilon& \textrm{\quad for $\bar c>2$},
\end{array} 
\right.
\ea
where we assumed that $y_*^{ij}$ is an anarchic matrix with complex coefficients all of the same order, and used the definition $\bar c=c_i+c_j$. The behavior for $\bar c>2$ reflects the fact that in this limit the 4D fermions become essentially fundamental (i.e. the mixing with the CFT operators becomes small) and the first term in~(\ref{mf}) dominates, see~\cite{TCC}. For $\bar c<2$ the mixing term in~(\ref{mf}) is the most important.

The strongest bound on the scale $\Lambda_\chi$ comes from rare $\Delta F=2$ processes, in particular $K\bar K$ mixing. An important correction to this latter process is generated by the 4-fermion contact term $Q_4^K\equiv(\bar d_R s_L)(\bar d_L s_R)$, where $d,s$ are fermion mass eigenstates and color indeces are contracted in the parenthesis. Imposing the bound Im$(C_4^K)\lesssim(1.6\times10^5$ TeV$)^{-2}$ on the coefficient $C_4^K$ of that operator [and assuming that all complex entries in the matrix $y_*^{ij}$ are $O(y_*)$] we find~(\ref{C4})
\ba\label{FCNCbound}
\Lambda_\chi> 1.6\times10^5\,\frac{g_*}{y_*}\frac{\sqrt{2m_dm_s}}{v\, r_{00}^H}\;\textrm{TeV}.
\ea
As a conservative approximation we assume that $r_{00}^H(\bar c)$ in~(\ref{FCNCbound}) is associated to the bottom quark (see Appendix~\ref{B} for more details). 

\begin{figure}
\begin{center}
\includegraphics[width=4.5in]{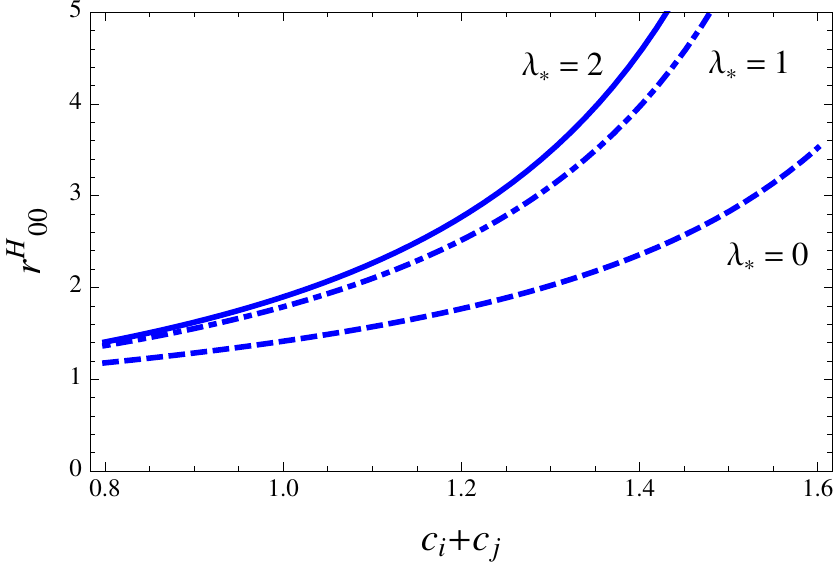}
\caption{Overlap function~(\ref{r00H}) for various values of the coupling $\lambda_*$ and $\epsilon=10^{-6}$ as a function of $\bar c=c_i+c_j$.\label{fr00H}}
\end{center}
\end{figure}

The constraint~(\ref{FCNCbound}) has formally the same structure as the one found for an IR-localized Higgs (see for instance~\cite{CFW}) except for the suppression induced by $r_{00}^H>1$. The very existence of this suppression is a consequence of the presence of the first contribution in~(\ref{mf}): the Yukawa couplings are enhanced for $\Delta<\infty$, namely for a bulk Higgs the overlap of the Higgs VEV with the bulk fermions is enhanced. For reference we plot $r_{00}^H(\bar c)$, defined in~(\ref{r00H}), as a function of $\bar c=c_i+c_j$ in fig.\ref{fr00H}.

For $\Lambda_F$ of the order the Planck mass one has $c_{bL}+c_{bR}\sim0.9-1.1$ and the suppression for $\lambda_*=0$ is close to $r_{00}^H\sim1.4$, in agreement with~\cite{Agashe}. With such a huge hierarchy we can gain up to a factor $r_{00}^H\sim2$ as compared to the IR localized Higgs case by taking $\lambda_*\sim2$. The bound found in~\cite{Gino} would then imply that KK masses as low as $m_\rho\sim4$ TeV are not excluded~\footnote{The estimate $m_\rho\sim4$ TeV is obtained from the results presented in~\cite{Gino} assuming that the constraint from CP violating effects in the $K^0\rightarrow2\pi$ system is unchanged, while taking into account the effect of the suppression coming from $r_{00}^H(\lambda_*=2)/r_{00}^H(\lambda_*=0)\sim1.5$.}.

For smaller hierarchies~\footnote{See~\cite{LRS} and~\cite{Bauer} for a discussion of FCNC bounds in models with relatively low hierarchies and $\Delta=\infty$.}, on the other hand, the bottom mass (say, matched at $\sim1$ TeV) is reproduced for larger values of the $c$'s, and hence a larger suppression (a larger $r_{00}^H$) is attained. Specifically, for a hierarchy of the order $\epsilon\sim10^{-6}$ we find $c_{bL}+c_{bR}\sim1.2-1.4$, which implies a suppression of the flavor bound by a factor $r_{00}^H=2-5$ as compared to the $\Delta\gg1$ Higgs case. For such relatively small hierarchies the bound on the KK scale becomes comparable to that implied by the EW fit, and a much richer phenomenology is predicted at energies accessible at the LHC.

A larger $y_*$ would further reduce the bound~(\ref{FCNCbound}), but at the same time it would spoil the perturbative expansion. A clear signal of this problem can be seen by observing that loop exchanges of KK (fermion) modes generate EDM-like operators with large coefficients $\propto y_*$~\cite{Agashe}\cite{Gino}. The effect reported here goes in the opposite direction: increasing the overlap function $r_{00}^H$ effectively enhances the Yukawa coupling of the light states, but at the same time reduces those of the heavy KK modes, with the net effect of \textit{improving} perturbation theory.

\section{Collider phenomenology}

The short distance phenomenology of the TCC scenario depends on the specific realization of the flavor sector. In this section we focus on physics at relatively low scales, say below $m_\rho$, at which the theory is basically described by the SM fermions and gauge fields, plus the composite Higgs. The phenomenology of this class of models has been studied in~\cite{SILH} and~\cite{GGS},~\cite{dilaton}.

Our low energy effective field theory is essentially controlled by the three free parameters of the model: $m_\rho$, defining the scale of the heavy KK modes, $m_\chi$, the Higgs/dilaton mass, and $f$, introduced in~(\ref{f}) and controlling the strength of the composite Higgs couplings. Using the theory of non-linear realization of the CFT one sees that the renormalizable couplings of the scalar to the SM are found by replacing those of a fundamental Higgs $h$ with~\cite{GGS}
\ba
\frac{h}{v}\rightarrow\frac{\bar\chi}{f}\left(1+O\left(\frac{m_\chi^2}{m_\rho^2}\right)\right),
\ea
where $\bar\chi$ is the canonically normalized dilaton field. Therefore, by varying the parameters $f$ and $m_\chi$ the model interpolates between a composite Higgs scenario (where the approximate symmetry is actually conformal invariance!) when $v\approx f$, and a RS scenario with a light dilaton and a heavy Higgs when $v<f$. In the former case the elastic $W^\pm W^\pm$ scattering will be perturbative up to the scale $m_\rho$, whereas in the latter case non-perturbative effects will set in not far from $1$ TeV.

While the renormalizable couplings between the composite Higgs and the SM fermions are formally similar to those of a fundamental Higgs boson, the production/decay into SM massless gauge fields is strongly enhanced because of the mixing between the gauge fields and the CFT composites. The dominant decay mode for the composite Higgs boson in the lower mass region (below the $W^\pm$ threshold) is by far $\chi\rightarrow gg$, with a branching ratio of the order $\textrm{BR}(\chi\rightarrow gg)\sim80\%$.

The composite Higgs $\chi$ also mediates FCNC processes at tree level~\cite{Toharia}~\cite{dilaton}. These are controlled by the dimensionful parameter $v/fm_\chi$ and are generally within the bounds for either $m_\chi$ not too small or $v< f$.

Another distinctive signature of the model is found by studying the composite Higgs potential. Expanding~(\ref{Ldil}) in the canonically normalized field we have
\ba
\frac{(\partial\bar\chi)^2}{2}-\Lambda_\chi^2v^2\sum_n \frac{g_n(\lambda_*)}{n\textrm{!}}\left(\frac{\bar\chi}{f}\right)^n.
\ea
The couplings $g_n$ increase in magnitude from $\lambda_*=0$, where they vanish, to $\lambda_*=2$, where they are maximal. Their value is $O(1)$ in the range $0<\lambda_*\lesssim1$, but significantly depart from unity for $1<\lambda_*\lesssim2$, where they scale as $g_n\sim2n\textrm{!}\lambda_*^{n-1}$. Such huge deviations from the values characterizing a weakly coupled Higgs sector may provide important signals at linear colliders.

\section{Conclusions}

We argued that a bulk Higgs with a mass close to the Breitenlohner-Freedman bound can naturally stabilize the size of the extra dimension at a hierarchically large length scale. There is no need for additional stabilizing fields in this class of theories: the Higgs does the job. Hierarchies of the order $\epsilon=10^{-6}$ are generated without fine-tuning.

Our construction predicts a Higgs VEV maximally spread in the bulk. This fact has important consequences for phenomenology.

We showed that the EW parameters are compatible with KK masses of the order $m_\rho\approx2$ TeV, and studied FCNC processes in a class of models that address the flavor problem of the SM. For very large hierarchies (unnatural in the present framework) our results agree with previous studies~\cite{Agashe}\cite{Gino}. For moderately small, but still phenomenologically viable hierarchies we found that the bound on $m_\rho$ from rare events is suppressed by $O(2-5)$ factors compared to the case of an IR-localized Higgs. In such a scenario the flavor bound on the new physics scale approaches that of the EW precision tests.

The Higgs boson in our framework emerges as a light pseudo-dilaton with renormalizable couplings to the SM fields naturally set by the weak scale. Radiative corrections to the Higgs mass are cutoff at the dynamical scale thanks to the gravitational redshift, and are therefore under control. The relatively low KK mass scale allowed by the experimental fit prevents the emergence of a little hierarchy problem, and implies a rich phenomenology at energies accessible at the LHC.

\acknowledgments
I thank Kaustubh Agashe for comments on the manuscript and Ian Shoemaker for discussions. This work has been supported by the U.S. Department of Energy at Los Alamos National Laboratory under Contract No. DE-AC52-06NA25396.

\appendix

\section{\label{A}The dilaton potential}

In this Appendix we derive the dilaton potential for the class of quadratic CFT deformations 
\ba\label{defA}
\delta{\cal L}=\lambda{\cal O}^\dagger{\cal O},
\ea
where ${\cal O}$ is a scalar operator of dimension $\Delta_+=\frac{d}{2}+\nu>\frac{d}{2}$ (the case of scaling dimension $\frac{d-2}{2}<\Delta_-=\frac{d}{2}-\nu<\frac{d}{2}$ can be analyzed in a similar fashion), and $d$ is the space-time dimension. We assume that ${\cal O}$ is charged under some flavor symmetry forbidding the linear term $J{\cal O}$. 


We work at quadratic order in the dual scalar $\phi$ and in a generic space-time dimension. Our lagrangian reads
\ba\label{app}
{\cal S}_\phi=\int d^dx\bigg\{\int_{1/\mu}^{1/\chi} dz\sqrt{-g}\left(|\nabla\phi|^2-m^2|\phi|^2\right)-\sqrt{-g_{UV}}\,V_{UV}-\sqrt{-g_{IR}}\,V_{IR}\bigg\},
\ea
with the IR potential forcing $\phi(1/\chi)=\phi_*$. In order to implement the deformation~(\ref{defA}) we choose the following UV potential~\cite{multi}
\ba
V_{UV}=\left(\Delta_+-\lambda(\mu)\right)\Lambda\,|\phi|^2
\ea
which induces the UV boundary condition $z\phi'=(\Delta_+-\lambda)\phi$ at the RG scale $z=1/\mu$. The solution of the equation of motion is 
\ba
\phi=\alpha z^{\Delta_-}+\beta z^{\Delta_+}, \quad\quad\textrm{and}\quad\quad\Delta_\pm=2\pm\nu=2\pm\sqrt{\frac{m^2}{\Lambda^2}+\frac{d^2}{4}}
\ea
whereas $\alpha,\beta$ are determined by the boundary conditions. The UV boundary condition can now be re-written as
\ba\label{coupling}
\lambda(\mu)=\frac{2\nu}{1+\frac{\beta}{\alpha}\mu^{-2\nu}}=2\nu\frac{\mu^{2\nu}}{\mu^{2\nu}+\left(\frac{2\nu-\lambda(\Lambda)}{\lambda(\Lambda)}\right)\Lambda^{2\nu}}.
\ea
Notice that the beta function of the coupling $\lambda$ is $\beta_\lambda=2\nu\lambda\left(1-\frac{\lambda}{2\nu}\right)$ as expected from field theory arguments~\cite{Vecchi}\cite{multi}.

The effective potential $V_{eff}$ for the dilaton $\chi$ at leading order in $\phi_*^2/M^3$ is given by the on-shell action
\ba
{\cal S}_\phi=\int d^dx\left[-V_{eff}\right],
\ea
and reads
\ba\label{VeffA}
V_{eff}=\frac{\chi^{d-1}}{\Lambda^{d-1}}\left[\phi\phi'\right]\Big\arrowvert_{z=1/\chi}=\chi^d\left[\Delta_+-\lambda(\chi)\right]\frac{\phi_*^2}{\Lambda^{d-1}}.
\ea
The effective potential is of the form $V_{eff}=\chi^dP(\lambda(\chi))$ required by RG invariance, see also~\cite{RZ}. Higher order terms in $\lambda(\chi)$ are expected to appear at next to leading order in the backreaction (see Appendix~\ref{A'}). Notice that the spurionic dependence on the cutoff in~(\ref{VeffA}) is renormalized away with the definition~(\ref{mW}).

From~(\ref{coupling}) one sees that the theory develops a dynamical scale in the phase $\lambda>2\nu$. The IR scale scale $\langle\chi\rangle=\Lambda_\chi$ is typically set by $\lambda(\Lambda_\chi)=O(1)>\lambda(\Lambda)$ [sometimes written as $\lambda(\Lambda_\chi)=\infty$] and it is given by $\Lambda_\chi^{2\nu}\sim\left(\frac{\lambda(\Lambda)-2\nu}{\lambda(\Lambda)}\right)\Lambda^{2\nu}$. For $2\nu<\lambda(\Lambda)$ and $2\nu$ somewhat smaller than 1, the vacuum condition naturally implies a hierarchical relation between IR and UV scales.

\section{\label{A'}The vacuum of the scalar/gravity system: backreaction included}

It is instructive to see what the vacuum configuration should look like in an exact solution of the scalar/gravity system~(\ref{tot}). 

Defining the function $W(\alpha)$ by $\beta=\frac{dW}{d\alpha}$, the vacuum configuration for a scalar/gravity system with $AdS_{d+1}$ boundary is determined by the absolute minimum of the potential~\cite{Tom}
\ba\label{W}
{\cal V}[\alpha]=W(\alpha)+s\frac{\Delta_-}{d}|\alpha|^{d/\Delta_-},
\ea
[where $s$ is a positive number that depends on the scalar potential] in the sense that given a boundary condition relating the coefficients $\alpha,\beta$ (i.e. given $W$) the quantity $\alpha$ (and hence $\beta$, and with it the full solution of the coupled equations of motion) is determined by the minimum of~(\ref{W}).

The effective potential~(\ref{W}) represents the total energy of the scalar/gravity system, and in particular it accounts for the backreaction of the scalar on the geometry. This is manifests in~(\ref{W}): the first term encodes the presence of CFT deformations [the second term in~(\ref{Veff})], the second term is proper of the CFT and it is thus scale invariant [this is the first term in~(\ref{Veff})].

Specializing on the deformation~(\ref{defA}) and using the boundary condition~(\ref{coupling}) we see that the theory defined in Appendix~\ref{A} has
\ba\label{W1}
{\cal V}=\left(\frac{2\nu-\lambda(\mu)}{\lambda(\mu)}\right)\mu^{2\nu}\frac{\alpha^2}{2}+s\frac{\Delta_-}{d}|\alpha|^{d/\Delta_-},
\ea
where we ignored the constant term [the constant of integration of $W$] that determines the $d$-dimensional cosmological constant [the last term in~(\ref{Veff})]. The minimum of ${\cal V}$ is thus found at the scale
\ba
\alpha=\mu^{\Delta_-}\left(\frac{\lambda-2\nu}{\lambda s}\right)^{\Delta_-/2\nu}\equiv\Lambda_\chi^{\Delta_-}.
\ea
This prediction is perfectly consistent with the field theory expectations: it has the correct mass dimension, it is RG invariant, and it exists only in the branch $\lambda>2\nu$; up to an irrelevant [$s$-dependent] constant the vacuum is therefore set by the condition $\lambda(\Lambda_\chi)=\infty$, as already anticipated.

The spontaneous breaking of the chiral symmetry [parametrized by $\beta\neq0$ in this example] is a consequence of the coupling becoming strong in the phase $\lambda>2\nu$. Whether the theory actually confines or not depends on the specific form of the bulk potential of the scalar [i.e. on the nature of the CFT]. In the realization presented in this paper confinement is assumed and mimicked by an IR boundary.

To further appreciate the result~(\ref{W}) it is useful to consider an explicit field theory example. We consider the Nambu Jona-Lasinio model in dimensions $2\leq d<4$, for which an exact planar solution for the effective potential is known. The model is defined by
\ba
{\cal L}_{\textrm{\tiny{NJL}}}&=&\bar\psi i\partial\psi+\frac{G}{2}(\bar\psi\psi)^2\\\no
&=&\bar\psi i\partial\psi-\frac{\alpha^2}{2G}+\bar\psi\psi\alpha
\ea
where in the last step we introduced the auxiliary field $\alpha$. This model belongs to the class of theories defined by the formal expression CFT +~(\ref{defA}). In the present case the CFT is trivial [it reduces to the fermion kinetic energy] and ${\cal O}=\bar\psi\psi$ has dimension $\Delta_+=d-1$ [this gives $\Delta_-=1$ in any dimension]. Integrating out the fermionic fields [assumed to be vectors of a global $U(N)$ symmetry] we find the effective potential ${\cal V}[\alpha]$ for the \textit{source} $\alpha$ [the integral is in Euclidean momentum space]:
\ba
\frac{\partial{\cal V}}{\partial\alpha}&=&\frac{\alpha}{G}-4\alpha\int\frac{d^dp}{(2\pi)^d}\frac{1}{p^2+\alpha^2}\\\no
&=&\frac{\alpha}{\bar G_*}\left[\frac{\bar G_*-\bar G}{\bar G}\mu^{d-2}+|\alpha|^{d-2}\right].
\ea
In the last equality we introduced the renormalized coupling $\bar G(\mu)$ and an arbitrary RG scale $\mu$ via the definition
\ba
\frac{\mu^{d-2}}{\bar G(\mu)}=\frac{1}{G}-\int\frac{d^dp}{(2\pi)^d}\frac{1}{p^2+\mu^2}.
\ea
The renormalized, dimensionless coupling satisfies the RG equation $\mu\frac{d\bar G}{d\mu}=(d-2)\bar G\left(1-\frac{\bar G}{\bar G_*}\right)$, with $\bar G_*$ a number [UV fixed point analog to $2\nu$ of Appendix~(\ref{A})]. One readily sees that the effective potential ${\cal V}[\alpha]$ of our field theory has precisely the form~(\ref{W1}), with a non-trivial vacuum obtained in the strong phase $\bar G>\bar G_*$ [analog to $\lambda>2\nu$ of our previous example]. The remarkable fact is that~(\ref{W}) holds even for non-trivial CFTs!

\section{\label{B}Yukawa texture: an example}

In this section we consider an explicit example of Yukawa texture arising from the general structure~(\ref{mf}) and derive an expression for the coefficient of the flavor violating operator in~(\ref{4ferm}). We focus on the down type quarks, for which the FCNC bounds are stronger.

For a hierarchy of order $\epsilon\sim10^{-6}$ and using~(\ref{mf'}) we see that the coefficients $c_i$ for the three generations ($i=1,2,3$) are typically $c_1\gtrsim1.5$, $c_2\lesssim1$, $c_3\sim0.5$. As a result the fermion mass matrix has the structure
\ba\label{texture}
m_f \sim m_3
\left(
\begin{array}{ccc}
\frac{m_1}{m_3} & \frac{m_1}{m_3} & \frac{m_1}{m_3} \\
\frac{m_1}{m_3} & \frac{m_2}{m_3} & \sqrt{\frac{m_2}{m_3}} \\
\frac{m_1}{m_3} & \sqrt{\frac{m_2}{m_3} } & 1
\end{array} \right),
\ea
with $m_1\approx y_*\frac{v}{\sqrt{2}}\epsilon\log1/\epsilon\gg y_*\frac{v}{\sqrt{2}}f_1^2$ and $m_{2,3}\approx y_*r_{00}^H\frac{v}{\sqrt{2}}f_{2,3}^2$. The functions $f_i$ are defined in~(\ref{f}) and $r_{00}^H$ in~(\ref{r00H}). In writing~(\ref{texture}) we assumed that the 5D Yukawa matrix is anarchic and with complex coefficients all of the order $y_*$, and ignored $O(1)$ numbers. 

The symmetric texture~(\ref{texture}) is diagonalized by a unitary matrix $D$ of the form [again ignoring factors $O(1)$]
\ba\label{D}
D\sim \left(
\begin{array}{ccc}
1 & \sqrt{\frac{m_2}{m_1}} & \sqrt{\frac{m_3}{m_1}} \\
\sqrt{\frac{m_1}{m_2}} & 1 & \sqrt{\frac{m_3}{m_2}} \\
\sqrt{\frac{m_1}{m_3}} & \sqrt{\frac{m_2}{m_3}} & 1
\end{array} \right).
\ea
The mass basis $\psi'_{L,R}$ is given in terms of the interaction basis $\psi_{L,R}$ by the relation $\psi_{L,R}^i=D_{ij}{\psi'}_{L,R}^j$. Notice that the matrix~(\ref{D}) has the very same texture found in the absence of the direct Higgs coupling in~(\ref{mf}), i.e. in the case $r_{00}^H=1$.

Let us now focus on chirality-changing 4-fermion operators involving the first two generations, and assume they are generated by the tree-level exchange of a flavor singlet vector [this assumption is justified in our model, where the dominant contribution to $\epsilon_K$ is induced by the tree-level integration of the KK gluon tower]. In this case the relevant contact term is
\ba
\frac{g_*^2}{\Lambda_\chi^2}\sum_if_i^2\psi_L^i\psi_L^i\sum_jf_j^2\psi_R^j\psi_R^j.
\ea
Projecting on the $1,2$ directions we find an expression for the coefficient $C_4^K$ introduced before~(\ref{FCNCbound}):
\ba\label{C4}
C_4^K&\approx&\frac{g_*^2}{\Lambda_\chi^2}\left[f_1^2D_{12}+f_2^2D_{21}+f_3^2D_{31}D_{32}\right]^2\\\no
&\approx&\frac{1}{\Lambda_\chi^2}\frac{g_*^2}{y_*^2}\frac{2m_1m_2}{v^2}\frac{1}{(r_{00}^H)^2}
\ea
where the first term in the square bracket is negligible compared to the remaining two [comparable in magnitude among each other] because of the UV dominance of the Yukawa matrix for $i=1$. In the last step we made use of the relation $m_{2,3}\approx y_*r_{00}^H\frac{v}{\sqrt{2}}f_{2,3}^2$. 

In estimating~(\ref{C4}) we ignored $O(1)$ factors coming from the non-universalities in the Yukawa matrix and the interference of the various terms in play. These simplifications are justified because our present aim is to compare the flavor bound in the TCC model with the estimate of~\cite{CFW} found in the case of an IR-localized Higgs [see the footnote~\ref{foot}]. The final expression for $C_4^K$ reveals that the main difference between these two scenarios is encoded in the factor $1/(r_{00}^H)^2\leq1$. A quantitative estimate of this suppression is given in the text.


 \end{document}